\documentstyle[12pt]{article}

\def\ie{\hbox{\it i.e.}{}}      
\def\eg{\hbox{\it e.g.}{}}      \def\cf{\hbox{\it cf.}{}}

\font\tenrm=cmr10
\font\tenit=cmti10
\font\tenbf=cmbx10
\font\elevenbf=cmbx10 scaled\magstep 1
\font\elevenrm=cmr10 scaled\magstep 1
\font\elevenit=cmti10 scaled\magstep 1
\font\ninebf=cmbx9

\font\nineit=cmti9

\begin{document}
\newcommand{\bibit}{\nineit}
\newcommand{\bibbf}{\ninebf}
\renewenvironment{thebibliography}[1]
{   \begin{list}{\arabic{enumi}.}
    {\usecounter{enumi} \setlength{\parsep}{0pt}
     \setlength{\itemsep}{2pt} \settowidth{\labelwidth}{#1.}
     \sloppy
    }}{\end{list}}

\parindent=3pc
\newcommand{\sbar}{\,\overline{\! S}}
\newcommand{\barz}{\,\overline{\! Z}}
\newcommand{\zbar}{\bar{z}}
\newcommand{\tbar}{\overline{T}}
\newcommand{\z}{\zeta}
\newcommand{\zb}{\bar{\zeta}}
\newcommand{\psibar}{\overline{\Psi}}
\newcommand{\cm}{{\elevenit Commun. Math. Phys.} }
\newcommand{\pr}{{\elevenit Phys. Rev.} }
\newcommand{\pl}{{\elevenit Phys. Lett.} }
\newcommand{\np}{{\elevenit Nucl. Phys.} }

\begin{titlepage}
\begin{flushright}
\hfill{CPTH-C218.0193}\\[1mm]
January 1993
\end{flushright}
\vskip 1cm

\begin{center}{
{\tenbf STRING LOOP CORRECTIONS TO GAUGE AND YUKAWA COUPLINGS\footnote{\rm
Based on talks presented at the International Workshop on String Theory,
Quantum
Gravity and the Unification of Fundamental Interactions, Roma, 21-26
September
1992, at the 7th Meeting of the American Physical Society, Division of
Particles
and Fields, Fermilab, 10-14 November 1992, and at the 26th Workshop: ``From
Superstrings to Supergravity", Erice, 5-12 December 1992.\\}}
\vglue 1.0cm
{\tenrm I. ANTONIADIS\\}
\baselineskip=13pt
{\tenit Centre de Physique Th\'eorique, Ecole Polytechnique\\}
\baselineskip=12pt
{\tenit  91128 Palaiseau, France\\}
\vglue 0.3cm
{\tenrm and}\\
\vglue 0.3cm
{\tenrm T.R. TAYLOR\\}
\baselineskip=13pt
{\tenit Department of Physics, Northeastern University\\}
\baselineskip=12pt
{\tenit  Boston, MA 02115, U.S.A.\\}
\vglue 0.8cm
{\tenrm ABSTRACT}}
\end{center}
\vglue 0.3cm
{\rightskip=3pc
 \leftskip=3pc
 \tenrm\baselineskip=12pt
 \noindent
We report on the recent progress in computing the effective supergravity
action
from superstring scattering amplitudes beyond the tree approximation. We
discuss the moduli-dependent string loop corrections to gauge,
gravitational and Yukawa couplings.
\vglue 0.6cm}
\end{titlepage}
\baselineskip=14pt
\vglue 0.6cm
{\elevenbf\noindent 1. Introduction}
\vglue 0.4cm
\elevenrm
The basic property of string theory, which makes it so attractive from the
point
of view of particle physics, is that the physical couplings and masses are
in
principle calculable. They are determined by the vacuum expectation values
(VEVs) of massless scalar fields, like the dilaton and moduli. The latter
are
scalar fields with flat potential, whose VEVs correspond to
compactification
parameters and determine the size and shape of the internal space. Any
serious attempt to compute the low-energy parameters from string theory
must
address two basic questions: 1) How do masses and couplings depend on the
VEVs
of dilaton, moduli, Higgs scalars {\em etc}.? and 2) What fixes these VEVs?
In
the past several years, there has been some steady progress towards
answering the second question, although the general perception is that the
problem of scalar VEVs still escapes a satisfactory solution. On the other
hand, there has been a lot of progress towards answering the first
question,
which we will generically call the problem of moduli dependence of physical
parameters.

A very efficient method of studying the moduli dependence of low-energy
parameters, developed in the last couple of years, relies on the
computation of
the effective supergravity action describing the physics of massless string
excitations. The moduli dependence of the effective action can be
determined by
evaluating the appropriate superstring scattering amplitudes.$^{1-5}$ In
this
process, supergravitational interactions are determined directly from
superstring theory. The moduli-dependent loop corrections obtained in this
way
give rise to the so-called threshold corrections to superstring unification
parameters,$^6$ and determine the boundary conditions of gauge and Yukawa
couplings at the string unification scale $M_{SU}$. They are also relevant
for
non-perturbative phenomena in string theory, which could generate
dynamically
potential for moduli providing a mechanism that fixes their VEVs. They may
also
be relevant for the mechanism of supersymmetry breaking, either via gaugino
condensation,$^7$ or with a large compactification radius.$^8$

The massless spectrum of any four-dimensional heterotic superstring model
contains the supergravity multiplet, gauge multiplets, and a large number
of
chiral multiplets. In addition, there is a dilaton which belongs to a very
distinct supersymmetry multiplet, together with the two-index antisymmetric
tensor -- the Kalb-Ramond field. The dilaton VEV plays the role of the
string
loop expansion parameter. Since the Kalb-Ramond field is equivalent to a
pseudoscalar axion, one usually represents the dilaton and its
supersymmetric
partners by one chiral multiplet $S$. The most general $N=1$ supergravity
action, describing local interactions involving up to two derivatives, is
characterized by three functions of chiral superfields:$^9$ the real
K\"ahler
potential $K$ which determines the kinetic terms, the analytic
superpotential
$W$ related to the Yukawa couplings, and the analytic gauge function $f$
associated with the gauge couplings.

At the tree level, the general stucture of the $f$-function and the
K\"ahler
potential is common to all compactifications:
\begin{equation}
f^{(0)}=kS~,~~~~~~~
K^{(0)}=-\ln (S+\sbar)+G^{(0)}(Z,\barz)\, ,
\label{tree}
\end{equation}
where $k$ is the integer level of the Ka\v{c}-Moody algebra that generates
the
gauge group, and $Z$ denote chiral superfields other than $S$. Thus, the
coupling constant $g_a$ of a gauge group factor $G_a$ is given at the
tree-level
by $g_a^2=g^2/k$, where $g$ is the four-dimensional string coupling. The
function $G^{(0)}(Z,\barz)$ as well as the superpotential $W(Z)$
depend on the details of compactification and can be determined by using a
number of different methods. At low energies, the most relevant
interactions
among matter and gauge fields correspond to effective operators of
dimension
four. Consequently, one may expand $G^{(0)}$ and $W$ up to quadratic and
cubic
order in matter fields, respectively, with moduli-dependent coefficients
which
one is interested to compute. These are known in many examples, in
particular
for the case of (2,2) compactifications which exhibit the property of
special
geometry implying that both functions are given in terms of a single
prepotential.$^{1,10}$
\vglue 0.6cm
{\elevenbf\noindent 2. Infrared divergences and threshold corrections}
\vglue 0.4cm
The tree-level effective action describes interactions of massless string
excitations at energies below the string scale. These include contact
interactions due to the propagation of heavy particles in
one-(massive)-particle
reducible diagrams. The masses of heavy particles depend on the moduli,
\eg\ the
radii of compactified dimensions, therefore the induced massless particle
interactions are also moduli-dependent. In order to compute the loop
corrections to the effective action, one should integrate all diagrams
involving
heavy particles propagating inside loops. In string theory, it is very
difficult
to separate heavy from massless particles in higher genus diagrams,
therefore
the computation of the effective action becomes more subtle than in field
theory. In fact integration over massless particles gives rise to infrared
divergences for on-shell amplitudes, associated with the running of low
energy
couplings.$^{6,11}$ In the analogous field-theoretical computations, such
logarithmic divergences are usually regulated by going off-shell, to
momentum
$\mu^2\neq 0$. It is very important to realize that in string theory, as
well as
in quantum field theory, the momentum-dependence of coupling constants is a
purely infrared effect, and therefore the corresponding $\beta$-function
coefficients of the $\mu^2\rightarrow 0$ divergence depend on the massless
particle content only.

Consider for instance the one-loop case. A generic on-shell amplitude $\cal
A$
corresponding to some physical coupling of the low-energy theory is written
as
an integral over the complex Teichm\"uller parameter $\tau =\tau_1
+i\tau_2$ of
the world-sheet torus inside its fundamental domain
$\Gamma\equiv\{|\tau_1|\leq
\frac{1}{2},~|\tau|\geq 1\}$:
\begin{equation}
{\cal A}=\int_{\Gamma}\frac{d^2\tau}{\tau_2}A(\tau,{\bar\tau}).
\label{ampl}
\end{equation}
The presence of massless particles propagating in the loop implies that the
integrand $A$ goes to a constant $A_0$ as $\tau_2\rightarrow\infty$, and
the
the integral over the Teichm\"uller parameter diverges in the infrared.
When the
logarithmic divergence is regularized and compared to the field-theoretical
$\overline{DR}$ scheme, it is converted to $\ln(M_{SU}^2/\mu^2)$, where
$\mu$ is
the infrared cutoff and $M_{SU}\simeq 5\times g \times 10^{17}$ GeV is the
string
unification scale.$^{6}$ The remaining finite part of the integral yields
the
moduli-dependent string threshold corrections:
\begin{equation}
{\cal A}=A_0\ln\frac{M_{SU}^2}{\mu^2} +
\int_{\Gamma}\frac{d^2\tau}{\tau_2}[A(\tau,{\bar\tau})-A_0].
\label{amplreg}
\end{equation}

When computing the moduli dependence of threshold corrections we use two
main
ingredients: 1) $N=1$ space-time supersymmetry to relate physical couplings
to
amplitudes involving pseudoscalars which receive contributions only from
odd
spin structures, and thus are much easier to compute. 2) The invariance of
the
superstring and its low energy effective theory under the discrete duality
group. This generalizes the simple $R\rightarrow 1/R$ duality, where $R$ is
the
radius of a circle on which one internal dimension is compactified to the
case
of more general compactifications. For instance, in the case of a plane one
can
define two complex parameters, each of them generating an $SL(2,Z)$ group
of
duality transformations $T\rightarrow 1/T$ and $T\rightarrow T+i$.
\vglue 0.6cm
{\elevenbf\noindent 3. Gauge couplings}
\vglue 0.4cm
The moduli dependence of threshold corrections to gauge couplings can be
determined in a way that circumvents the problem of infrared divergences.
It is
based on the observation that the $\beta$-function which is the coefficient
of
the infrared divergence is moduli independent and drops when one considers
derivatives with respect to the moduli. Consider for instance the 3-point
scattering amplitude involving one modulus $T$ and two gauge bosons. As a
consequence of supersymmetry, the $C\! P$-even and the $C\! P$-odd part of
this
amplitude are proportional,$^2$ implying:
\begin{equation}
\partial_T g^{-2}(T,\overline{T})=i\Theta_T\ ,
\label{thetat}
\end{equation}
where $\Theta_T$ is the axionic coupling of the pseudoscalar component of
$T$ to
gauge bosons. If $\Theta_T$ is the derivative with respect to $T$ of a
$\Theta$-angle which is a function of moduli, Eq.~(\ref{thetat}) implies
that
gauge couplings and the corresponding $\Theta$-angles are obtained as the
real
and the imaginary part, respectively, of analytic gauge functions $f$,
consistently with the general form of $N=1$ supergravity.$^9$ However, due
to
the integration over massless particles in the physical amplitude,
$\Theta_T$
may not in general be integrable and a useful relation to examine is the
integrability condition:
\begin{equation}
\partial_T\partial_{\tbar}g^{-2}(T,\tbar)=\frac{i}{2}
(\partial_{\tbar}\Theta_T-\partial_T\Theta_{\tbar})\
{\mathop=^?}\ 0\ .
\label{integr}
\end{equation}

Explicit one loop calculation in orbifolds shows that the integrability
condition Eq.~(\ref{integr}) is indeed violated and the full moduli
dependence
of threshold corrections to gauge couplings can be computed.$^2$ At higher
loops,
this dependence satisfies a non-renormalization theorem implying that the
one-loop result is exact.$^3$ For general compactifications, a general
formula
can be derived by integrating the one-loop expression of the axionic
coupling
$\Theta_T$ in Eq.~(\ref{thetat}):$^{4,5}$
\begin{equation}
\frac{1}{g^2}~=~-\frac{i}{32\pi^2}\int_{\Gamma}\frac{d^2\tau}{\tau_2}
\bar{\eta}^{-2} {\rm Tr}_R F(-1)^F (Q^2 -\frac{k}{4\pi\tau_2}) ,
\label{gauge}
\end{equation}
where $\eta$ is the Dedekind eta function, $Q$ is the gauge group
generator, and
the trace is over the Ramond sector of the internal $N=2$ superconformal
theory
with $U(1)$-charge operator $F$. Note that the integral in
Eq.~(\ref{gauge}) is
infrared divergent. The coefficient of the logarithmic divergence
$\frac{d\tau_2}{\tau_2}$ is:
\begin {equation}
b=\frac{1}{32\pi^2}(-3{\rm Tr}Q_V^2 + {\rm Tr}Q_M^2),
\label{betaf}
\end{equation}
where the two terms are the contribution of gauginos and matter fermions
with
$U(1)$-charges $F=\pm 3/2$ and $F=\pm 1/2$, respectively. The expression of
$b$
in Eq.~(\ref{betaf}) coincides with the field-theoretical one-loop
$\beta$-function of gauge couplings in $N=1$ supersymmetric Yang-Mills
theory.
It is worth noting that the same quantity ${\rm Tr}F(-1)^F$ was studied in
the
massive case, where $F = F_L - F_R$, as a new kind of ``index" for $N=2$
theories which depends only on chiral deformations.$^{12}$
\vglue 0.6cm
{\elevenbf\noindent 4. Gravitational couplings}
\vglue 0.4cm
The above results for gauge couplings are extended for the case of
gravitational
couplings.$^4$ In this case, there is no moduli dependent correction
to the Planck mass, at least up to the one loop order, and the role of
gauge
couplings and $\Theta$-angles is played by the coefficients of the
Gauss-Bonnet
combination and the $R\tilde{R}$ term, respectively. Eq.~(\ref{gauge}) is
then
valid for the gravitational coupling after replacing the gauge group
generator
$Q^2$ with
\begin {equation}
Q_{\rm grav}^2\equiv -\frac{1}{i\pi}\partial_{\bar{\tau}}\ln
(\bar{\eta}^2).
\label{qgrav}
\end{equation}
The infrared divergence can now be identify with the four-dimensional
one-loop
trace anomaly:
\begin {equation}
b_{\rm grav}=\frac{1}{32\pi^2}
[\frac{1}{6}(-3N_V+N_S)-\frac{11}{3}(-3+N_{3/2})],
\label{tranom}
\end{equation}
where $N_S$, $N_V$ and $N_{3/2}$ denote the number of chiral, vector and
spin-3/2 (massless) supermultiplets, while the first term in the second
bracket
accounts for the contribution of graviton and dilaton supermultiplets. This
expression agrees with the field-theoretical calculations of trace anomaly
coefficients, where the antisymmetric tensor contribution is {\em
different}
from that of a scalar.$^{13}$
\vglue 0.6cm
{\elevenbf\noindent 5. Integrability condition and duality anomalies}
\vglue 0.4cm
Going back to the integrability condition (\ref{integr}), using the
expression
of Eq.~(\ref{gauge}) one finds:$^{4,5}$
\begin{eqnarray}
\partial_T\partial_{\tbar}g^{-2}(T,\tbar) &=&
\frac{-i}{2{(2\pi)}^4}\int_{\Gamma}d^2\tau \partial_{\tau} \int{d^2\z}\,
\bar{\eta}^{-2} \langle Q^2 \Psi_T(\z)\psibar_{\tbar}(0)
\rangle_{\rm odd}\nonumber\\ & & \hspace{5mm}+
\frac{k}{8{(2\pi)}^5}\int_{\Gamma}\frac{d^2\tau}{{\tau_2}^2} \int d^2{\z}
\,\bar{\eta}^{-2}\, \langle\Psi_T(\z)\psibar_{\tbar}(0)
\rangle_{\rm odd}\, ,
\label{intexpr}
\end{eqnarray}
where $\Psi_T$ ($\psibar_{\tbar}$) are the corresponding chiral
(anti-chiral)
primary fields of the underlying $N=2$ internal superconformal theory with
dimension ($\frac{1}{2}$,1). Eq.~(\ref{intexpr}) gives the non-harmonicity
of
gauge couplings and contains two parts:
\begin{enumerate}
\item
The group-dependent part proportional to $Q^2$ is a total derivative in
$\tau$
and its contribution to the integral comes only from the boundary of the
moduli
space, namely the degeneration limit $\tau_2 \rightarrow \infty$. It is
therefore determined only from massless particles. In fact, all massless
chiral
fermions couple off-shell with the pseudoscalars$^9$ and generate one-loop
anomalous couplings of pseudoscalars to gauge bosons or gravitons which are
non-local. A detailed analysis of this term$^4$ was shown to reproduce the
field
theory computation of the anomalous graphs.$^{14,15}$
\item
The universal part ({\em i.e}.\ the term proportional to $k$) can be
identified
with the one-loop correction $G^{(1)}_{T\tbar}$ to the K\"ahler metric. The
corresponding one-loop contribution to the K\"ahler potential is:
\begin {equation}
K^{(1)}=-\ln [1-\frac{2}{(S+\sbar)}G^{(1)}(Z,\barz)]\, .
\label{k1loop}
\end{equation}
$G^{(1)}(T,\tbar)$ gives rise to one-loop kinetic terms that mix the moduli
$T$
with the dilaton $S$ which then couples universally to gauge bosons and
gravitons at the tree-level. This is usually called the Green-Schwarz term
because it can be interpreted as the compactification of the
ten-dimensional
term involved in the Green-Schwarz anomaly cancellation mechanism.$^{14}$
\end{enumerate}
The non-harmonicity of gauge couplings is also related to duality
anomalies.$^{14,15}$ In fact the first term in Eq.~(\ref{intexpr}),
associated
with one-loop anomalous graphs, generates anomalies in the duality
transformations at the effective field theory level. These are partially
cancelled by the contribution of the massive string modes which produce
{\em
local} one-loop corrections to the gauge kinetic functions $f^{(1)}(T)$,
obtained after integrating Eq.~(\ref{intexpr}) to Eq.~(\ref{gauge}). There
remains a gauge group-independent anomaly cancelled by the second term of
Eq.
(\ref{intexpr}), which plays a similar role as the Green-Schwarz term in
the
cancellation of ten-dimensional gauge anomalies.
\vglue 0.6cm
{\elevenbf\noindent 6. One loop K\"ahler metric and Yukawa couplings}
\vglue 0.4cm
As a consequence of the supersymmetric non-renormalization theorems in the
background field method, the superpotential does not receive any loop
corrections. However, the physical Yukawa couplings defined by the
fermion-scalar-fermion scattering amplitudes may receive loop corrections
which
arise from the wave function renormalization factors, \ie\ from the
corrections
to the K\"ahler metric. The computation described in the previous Sections
3-5
remains valid if one replaces the modulus $T$ with any matter field which
is
singlet with respect to the gauge group under consideration. Thus, the
one-loop
correction to the K\"ahler metric is given by the second term of
Eq.~(\ref{intexpr}) and corresponds to the variation of a K\"ahler
potential
given by the  second term  of Eq.~(\ref{gauge}):
\begin{equation}
G^{(1)}~=~\frac{i}{16{(2\pi)}^3}
\int_{\Gamma}\frac{d^2\tau}
{\tau_2^2}\bar{\eta}^{-2} {\rm Tr}_R F(-1)^F\ .
\label{g1loop}
\end{equation}
This can be checked independently by an alternative computation of the
one-loop
three-point amplitude involving two complex scalars and the antisymmetric
tensor field.$^{5}$ The result is:
\begin{equation}
G^{(1)}_{i\bar{\jmath}}
=\frac{1}{8{(2\pi)}^5}\int_{\Gamma}\frac{d^2\tau}{{\tau_2}^2} \int
d^2{\z} \,\bar{\eta}(\bar{\tau})^{-2}\,
\langle\Psi_i(\z)\psibar_{\bar{\jmath}}(0) \rangle_{\rm odd}\, ,
\label{gzz}
\end{equation}
in agreement with the field-theoretical expression from
Eq.~(\ref{intexpr}).

In the case of matter metric, the integral over the Teichm\"uller parameter
$\tau$ in Eq.~(\ref{gzz}) is infrared divergent. As in the case of gauge
couplings, these divergences are due to massless particles propagating in
the
loop. The coefficients of divergent terms correspond to the one-loop
anomalous
dimensions:$^5$
\begin{equation}
\gamma_i = -\frac{1}{64\pi^2}\{ -\frac{4}{k} C_2(R_i) +
\frac{1}{g^2}\sum_{j,k}|\lambda_{ijk}|^2 \} ,
\label{gammaf}
\end{equation}
where $C_2(R_i)$ is the quadratic Casimir of the representation $R_i$ to
which
the $i$-th field belongs, and $\lambda_{ijk}$ are the physical Yukawa
couplings.
The comparison with the field-theoretical anomalous dimensions shows that the
string computation implicitly uses a gauge in which the superpotential
remains unrenormalized. Again as in the case of gauge couplings, the
momentum-dependence of the physical Yukawa couplings in string theory turns
out
to be determined by the corresponding field-theoretical
$\beta$-functions.$^{16}$ The remaining finite part of
$G^{(1)}_{i\bar{\jmath}}$
gives the string  threshold corrections to wave function factors. These
corrections determine the boundary conditions for the physical Yukawa
couplings
$\lambda_{ijk}$ at the unification scale:
\begin{equation}
\lambda_{ijk}(M_{SU})=\lambda_{ijk}^{\rm tree}\, [1+g^2\,
(Y_i+Y_j+Y_k)]^{-1/2},
\label{yukawa}
\end{equation}
where $Y_i$ is defined as the finite part of
$G^{(1)}_{i\bar{\imath}}/G^{(0)}_{i\bar{\imath}}$.
\vglue 0.6cm
{\elevenbf\noindent 7. Explicit examples in orbifold models}
\vglue 0.4cm
As an example we will consider the symmetric orbifold models. These are a
particular case of $(2,2)$ compactifications, which possess $N=2$
world-sheet supersymmetry in both left and right moving sectors. The gauge
group is $E_8\otimes E_6\otimes H_2$, with all Ka\v{c}-Moody levels $k=1$
and
with $H_2$ a model dependent factor of rank-2. The relevant matter fields
transform as $27$ or $\overline{27}$ under $E_6$ and they are in one-to-one
correspondence with the moduli : $27$'s are related to $(1,1)$ moduli and
$\overline{27}$'s to $(1,2)$ moduli. Furthermore, the moduli metric is
block-diagonal with respect to these two types of moduli. An interesting
consequence of the right-moving $N=2$ tree-level Ward-identities is the
property of special geometry, which relates the tree-level moduli metric to
the
Yukawa couplings.$^{1,10}$

Here, we mainly focus our attention on the case of the three untwisted
moduli
$T_j$, $j=1,2,3$, describing the size of the three internal compactified
planes, which are in one-to-one correspondence with three untwisted
families
$A_j$. In this case, the superpotential and the tree-level K\"ahler
potential
are:
\begin{equation}
W=w_{123}A_1A_2A_3~,~~~~~~G^{(0)}=-\sum_{j=1}^3\ln
(T_j+\overline{T}_{\bar{\jmath}} - A_j\bar{A}_{\bar{\jmath}}) \ ,
\label{extree}
\end{equation}
where $w_{123}$ are numerical constants. The tree-level physical Yukawa
couplings are then $\lambda_{123}^{\rm tree}=\frac{g}{\sqrt 2}w_{123}$.

At the one-loop level, one has to sum over all sectors of boundary
conditions
of the orbifold group. The untwisted sector, which preserves $N=4$
space-time
supersymmetry, gives vanishing contribution to all $\beta$-functions and
threshold corrections. The twisted sector preserving only $N=1$
supersymmetry
leads to moduli-independent corrections, since the masses of the
corresponding
string states do not depend on the geometry of the compactified space.
Therefore,
the gauge and untwisted Yukawa couplings receive moduli-dependent
corrections
only from sectors that preserve $N=2$ space-time supersymmetry, which
appear
when one of the three internal planes remains invariant under the boundary
conditions. The variation $\partial_T\partial_{\tbar}g^{-2}_a$ can be
easily
evaluated from Eq.~(\ref{intexpr}) with the result:$^{2,3}$
\begin{equation}
\partial_T\partial_{\tbar}g^{-2}_a = {\tilde b}_a G^{(0)}_{T\tbar}~,~~~~~~
G^{(0)}_{T\tbar}=\frac{1}{(T+\tbar)^2} \ ,
\label{exint}
\end{equation}
where the coefficient ${\tilde b}_a$ is proportional to the
$\beta$-function of
the gauge group factor $G_a$ in the corresponding $N=2$ theory. It can be
writen
as a sum of two terms,
\begin{equation}
{\tilde b}_a = {\tilde b}_a^{\rm FT}+{\tilde b}^{\rm GS}\ ,
\label{bhat}
\end{equation}
where ${\tilde b}_a^{\rm FT}$ is the field-theoretical group-dependent
contribution and ${\tilde b}^{\rm GS}$ is the universal Green-Schwarz term,
in correspondence with the two terms of Eq.~(\ref{intexpr}). Explicit
evaluation$^5$ of the second integral shows that ${\tilde b}^{\rm GS}$ does
indeed subtract the contribution of the $N=1$ sector from the
field-theoretical
coefficient, so that ${\tilde b}_a$ is given entirely by $N=2$ sectors. The
same equation (\ref{exint}) is also valid for the gravitational couplings
with a
coefficient ${\tilde b}_{\rm grav}$ proportional to the trace anomaly of
the
corresponding $N=2$ theory.$^4$

After integrating the differential equation (\ref{exint}) using the
$SL(2,Z)$
duality symmetry one finds:
\begin{eqnarray}
\frac{1}{g^2_a(M_{SU})} &=& \frac{1}{g^2}-\tilde{b}_a\ln
[|\eta(iT)|^4(T+\tbar)]+c_a \nonumber\\ &+& \frac{B\overline{B}}{T+\tbar}
[\frac{1}{2}C_2(G_a)+{\cal{T}}(27)(1+\frac{\chi}{6}
-\frac{2}{g^2}\sum_{k,l}|\lambda_{Bkl}|^2)]+\dots \, ,
\label{exgauge}
\end{eqnarray}
where $c_a$ are moduli-independent constants and we also included the
contribution of the blowing-up moduli $B$ at the lowest order.$^4$ The
latter
deform orbifolds to smooth Calabi-Yau manifolds and are in one-to-one
correspondence with twisted families. $\chi$ is the number of generations
given
by the Euler number, and ${\cal T}(27)$ is the Dynkin index of the matter
representation which equals 3 and 0 for $E_6$ and $E_8$, respectively.

The one-loop K\"ahler metric and threshold corrections to Yukawa couplings
can
be explicitly computed from Eqs.~(\ref{gzz}) and (\ref{yukawa}). For the
untwisted moduli one obtains $G^{(1)}_{T\tbar}={\tilde b}^{\rm GS}
G^{(0)}_{T\tbar}$, where ${\tilde b}^{\rm GS}$ was defined in
Eq.~(\ref{bhat}).
A consequence of this multiplicative finite renormalization of the moduli
metric
is that special geometry is in general violated beyond the tree
approximation.
This is expected since from field-theoretical point of view special
geometry
is a consequence of $N=2$ space-time supersymmetry and ${\tilde b}^{\rm
GS}$ is
non vanishing only in $N=1$ sectors.$^5$

The wave function threshold corrections of an untwisted field $A_j$ (27 or
$\overline{27}$) associated with the $j$-th plane depends only on its
moduli
$T_j$, and {\em not\/} on the moduli of other planes:$^5$
\begin{equation}
Y_j=2\tilde{\gamma}_j\ln [|\eta (iT_j)|^4(T_j+\tbar_j)]+y_i\, ,
\label{exy}
\end{equation}
where $y_j$ is a moduli-independent constant and the coefficient
$\tilde{\gamma}_j$ is the anomalous dimension of the $A_j$-field in the
corresponding $N=2$ supersymmetric theory. Since this field belongs to an
$N=2$
vector supermultiplet, one has $\tilde{\gamma}_j=-\tilde{b}_j/2$, where
$\tilde{b}_j/2$ is the corresponding beta function coefficient of any gauge
subgroup that transforms $A_j$ non-trivially in the embedding $N=2$ theory.
The threshold corrections to Yukawa couplings can be computed by
substituting
Eq.~(\ref{exy}) into Eq.~(\ref{yukawa}). The boundary conditions for the
one-loop untwisted Yukawa couplings are then given by:
\begin{equation}
\lambda_{123}(M_{SU})=\frac{g_{E_6}(M_{SU})}{\sqrt{2}}w_{123}\,
[1+g^2_{E_6}(M_{SU})y_{123}]^{-1/2},
\label{exyuka}
\end{equation}
where $y_{123}$ are moduli-independent constants, and $g_{E_6}(M_{SU})$ is
the
one-loop $E_6$ gauge coupling at the unification scale, \cf\
Eq.~(\ref{exgauge}):
\begin{equation}
\frac{1}{g^2_{E_6}(M_{SU})} = \frac{1}{g^2}-\sum_{j=1}^3\tilde{b}^j_{E_6}
\ln [|\eta(iT_j)|^4(T_j+ \tbar_{\bar{\jmath}})]+c_{E_6}\, .
\label{exbound}
\end{equation}
As a result, the boundary relation between the untwisted Yukawa couplings
and
the $E_6$ gauge coupling at the unification scale does not receive any
moduli-dependent corrections at the one-loop level.
\vglue 0.6cm
{\elevenbf\noindent 8. Conclusions \hfil}
\vglue 0.4cm
Superstring theory provides a unique example of a perfectly consistent
unification of gauge and gravitational interactions, within the framework
of
local supersymmetry. The efforts described here have been motivated by the
desire to understand the low-energy limit of such a consistent theory. As a
result, we have now a very good understanding of the low-energy physics of
supergravity theory that describes not only the classical limit of string
theory, but also some interesting string loop effects, in particular the
threshold corrections. From the three functions which determine the
effective
$N=1$ supergravity, the two analytic ones are subject to non-renormalization
theorems. The superpotential is given entirely at the tree-level, while the
gauge kinetic function at the one loop. On the other hand, the one-loop
correction to the real K\"ahler potential is closely related to the new
``index" of $N=2$ superconformal theories. The infrared divergences were
identified with the field-theoretical $\beta$-functions of gauge and Yukawa
couplings, while the finite part define the moduli-dependent threshold
corrections which are calculable in any given model. Here, we presented
explicit examples in the case of orbifolds. In addition to possible
applications for low-energy physics, this ``bottom--up" approach to
superstring theory may well provide some new insights into the Planck-scale
physics.
\vglue 0.6cm
{\bf \noindent Acknowledgements \hfil}
\vglue 0.4cm
This work was supported in part by the Northeastern University Research and
Scholarship Development Fund, in part by the National Science Foundation
under
grant PHY-91-07809, in part by the EEC contracts SC1-0394-C, SC1-915053
and SC1-CT92-0792, and in part by a CNRS-NSF collaborative grant.
\vglue 0.6cm
{\bf \noindent References \hfil}
\vglue 0.4cm

\end{document}